\documentclass[aps,prl,twocolumn,showpacs,superscriptaddress,groupedaddress,reprint]{revtex4-1}
\usepackage{graphicx}
\usepackage{mathrsfs}
\usepackage{natbib}
\usepackage{color}
\begin{document}

\widetext
\title{Fluctuation-stabilized marginal networks and anomalous entropic
  elasticity}

\author{M. Dennison$^{1}$} 
\author{M. Sheinman$^{1}$} 
\author{C. Storm$^{2}$} 
\author{F.C. MacKintosh$^{1}$} 

\affiliation{$^{1}$Department of Physics and Astronomy,
  VU University, De Boelelaan 1081, 1081 HV Amsterdam, The
  Netherlands}
\affiliation{$^{2}$Department of Applied Physics and Institute for Complex Molecular Systems, Eindhoven University of Technology,
  P.O. Box 513, NL-5600 MB Eindhoven, The Netherlands}

\date{\today}

\begin{abstract}
We study the elastic properties of thermal networks of Hookean
springs. In the purely mechanical limit, such systems are known to
have vanishing rigidity when their connectivity falls below a
critical, isostatic value. In this work we show that thermal networks
exhibit a non-zero shear modulus $G$ well below the isostatic point,
and that this modulus exhibits an anomalous, sublinear dependence on
temperature $T$. At the isostatic point, $G$ increases as the
square-root of $T$, while we find $G\propto T^\alpha$ below the
isostatic point, where $\alpha\simeq0.8$. We show that this anomalous
$T$ dependence is entropic in origin.
\end{abstract}
\maketitle

The stiffness of elastic networks depends on the mechanical properties
of their constituents as well as their connectivity, which can be
measured by the average coordination of nodes. Maxwell showed that a
network of simple springs will only become rigid once the connectivity
exceeds a critical, \emph{isostatic} value at which the number of
constraints just balances the number of internal degrees of freedom
\cite{ref:Maxwell}. This purely mechanical argument can be used to
understand the rigidity of such diverse systems as amorphous solids
\cite{ref:Lubensky_as}, jammed particle packings and emulsions
\cite{ref:Liu,ref:Hecke} and even some folded proteins
\cite{ref:Rader_pf}. Interestingly, under-constrained systems that are
mechanically \emph{floppy} can become rigid when thermal effects are
present. Perhaps the best known example of this is \emph{entropic}
elasticity of flexible polymers \cite{ref:deGennes}. Even a single,
freely-jointed chain that is mechanically entirely floppy becomes
elastic at finite temperature $T$: such chains resist extension with a
spring constant that is proportional to $T$. At the level of networks
of such chains, the macroscopic shear modulus also grows proportional
to $T$ \cite{ref:deGennes,ref:Alexander}. Many systems, including
network glasses \cite{ref:Thorpe_ng,ref:Thorpe_ng2,ref:Phillips} and
some biopolymer networks \cite{ref:Storm,ref:Frey,ref:Weitz} can be
considered intermediate between a purely mechanical regime well above
the isostatic point, and a purely thermal or entropic regime below the
isostatic point. However, very little is known about thermal effects
of such systems near the isostatic point
\cite{ref:Rubinstein,ref:Barriere,ref:Plischke,ref:Tessier}.

\begin{figure}[b]
  \begin{center}
      \includegraphics[width=1.0\columnwidth]{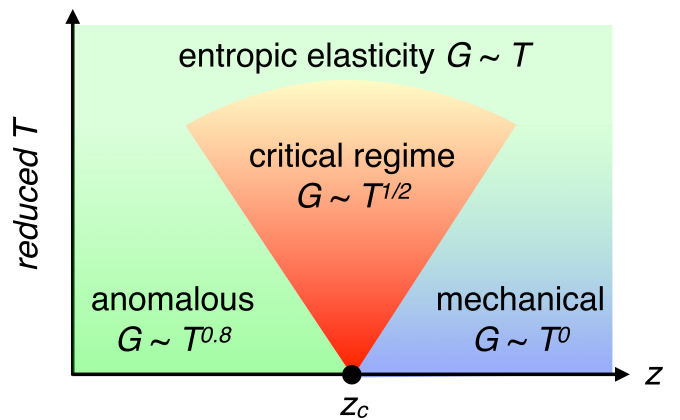}
    \caption{Schematic phase diagram of thermal networks in the $T-z$
      representation, where `reduced $T$' is the ratio of the
      temperature to the spring energy and $z$ is the connectivity,
      with critical connectivity $z_{\mathrm{c}}$. Reminiscent of
      quantum critical points \cite{ref:Sachdev,ref:Coleman}, we find
      a critical regime that broadens out for temperatures above the
      $T=0$ critical point.}
   \label{fig:PD}
  \end{center}
\end{figure}

Here we show that simple model networks, consisting of randomly
diluted springs, can be stabilized by thermal fluctuations, even at
low connectivity for which they would be floppy at zero
temperature. Interestingly, we find that the linear shear modulus $G$
exhibits anomalous temperature dependence both at and below the
isostatic point. Specifically, we find that $G\propto T^\alpha$, where
$\alpha<1$. This is surprising since one might have expected, in
analogy with freely-jointed chains, that such networks would exhibit
ordinary entropic elasticity ($G\propto T$) below the isostatic point,
as the mechanically floppy modes are excited thermally. Moreover, we
find two distinct anomalous entropic elasticity regimes in the
connectivity-temperature phase diagram, with the Maxwell isostatic
point acting as a zero-temperature critical point (Fig.~\ref{fig:PD}).

\begin{figure*}[!t]
  \begin{center}
      \includegraphics[height=2.0\columnwidth,angle=270]{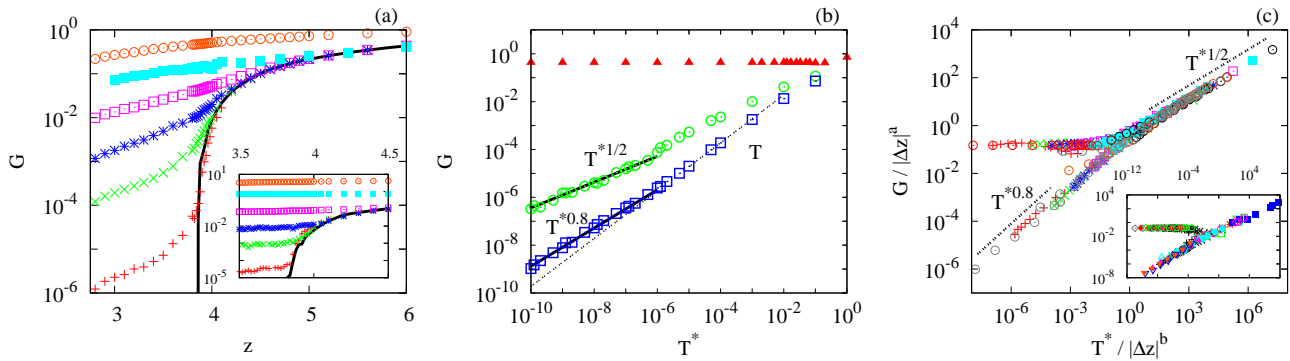}
    \caption{The network shear modulus $G$ in units of
      $k_{\mathrm{sp}}$ for $N = 3600$ nodes connected by phantom
      springs of rest length $\ell_{0}=1$. The main plots are for
      fixed area $A = A_{0}$, the area of a relaxed, fully connected
      network at $T = 0$. The corresponding results for self-avoiding
      springs at $P = 0$ are shown in the insets. (a) $G$ as a
      function of $z$ for $T^{*} =
      k_{B}T/k_{\mathrm{sp}}\ell^{2}_{0}=10^{-6}$ (lower), $10^{-4}$,
      $10^{-3}$, $10^{-2}$, $10^{-1}$ and $1$ (upper).  Solid line
      shows $T=0$ results. (b) $G$ as a function of $T^{*}$ for $z=6$
      (triangles), $z\simeq z_{\mathrm{c}}=3.857$ (circles) and $z=3$
      (squares). (c) Scaling of the shear modulus using the form $G =
      k_{\mathrm{sp}}|\Delta z|^{a} \mathscr{F}\left(T^{*}|\Delta
      z|^{-b}\right)$, where $\Delta z = z-z_{\mathrm{c}}$, for
      $T^{*}<10^{-5}$. The two branches on the left hand side
      correspond to $z>z_{c}$ (upper) and $z<z_{c}$ (lower). In both
      systems, the asymptotes and exponents ($a=1.4$ and $b=2.8$) are
      the same.}
   \label{fig:G}
  \end{center}
\end{figure*}

We perform Monte Carlo (MC) simulations on 2D spring networks that
consist of $N=n^{2}$ nodes, arranged on a triangular lattice, that are
connected by $N_{\mathrm{sp}}=zN/2$ springs, where $z$ is the average
connectivity ($z=6$ for the fully connected network). Periodic
boundary conditions are used in all directions. To avoid network
collapse \cite{ref:Boal}, we consider two cases: one in which we keep
the system area $A$ fixed and treat the springs as `phantom' (i.e., we
ignore steric interactions, and hence the springs are potentially
overlapping), and one where we fix the system pressure $P$ and prevent
the springs from overlapping (self-avoiding springs). In both cases
the system energy is given by
\begin{equation}
\label{eq:E_sp}
{\cal U} = \frac{k_{\mathrm{sp}}}{2}\sum^{N_{\mathrm{sp}}}_{i=1}\left(\ell_{i}-\ell_{0}\right)^{2},
\end{equation}
where $\ell_{i}$ is the length of spring $i$, $\ell_{0}$ is the rest
length and $k_{\mathrm{sp}}$ is the spring constant. In order to lower
the connectivity of the system we set $k_{\mathrm{sp}}=0$ for randomly
chosen springs. For the phantom network this is identical to removing
springs, while for the self-avoiding network this method has the
advantage of computational efficiency over simply removing the
springs, since springs with $k_{\mathrm{sp}}=0$ still contribute
steric interactions and hence the nodes are essentially confined to a
`cell' by the surrounding springs.

To find the critical (\emph{isostatic}) point $z_{\mathrm{c}}$, for
the onset of rigidity at $T=0$, we use a conjugate gradient algorithm
to calculate $G$. For 2D networks $z_{\mathrm{c}}\simeq4$
\cite{ref:Maxwell,ref:Jacobs}, although due to finite size effects
this will be somewhat smaller for each $N$ value studied
\cite{ref:Chase_bend}. We then increase $T$ in steps and allow the
systems to equilibrate using MC simulations, obtaining configurations
under shear. We note that there is an additional critical point
$z_{\mathrm{P}}\simeq2.084$ \cite{ref:Percolation}, corresponding to
the connectivity percolation threshold, below which there is no
connected path through the network. For $T>0$ the shear modulus is
finite between $z_{\mathrm{p}}$ and $z_{\mathrm{c}}$
\cite{ref:Plischke}.

In order to shear the systems, we use Lees-Edwards boundary conditions
\cite{ref:Lees_Edwards} to apply a shear strain $\gamma$. The shear
modulus $G$ is then given by
\begin{equation}
\label{eq:G_shear}
G =\frac{1}{A}\frac{\partial ^{2} {\cal F}}{\partial\gamma^{2}},
\end{equation}
where ${\cal F}$ is the free energy of the system. It is not possible
to directly calculate ${\cal F}$ from MC simulations, so we calculate
the linear shear modulus $G$ as described in \cite{ref:Hoover,ref:SM}.
Moreover, since $G$ has units of $k_{\mathrm{sp}}$ in $2D$, we express
$G$ throughout in units of $k_{\mathrm{sp}}$.

At low temperatures we find that the shear modulus closely follows the
zero-temperature behavior, decreasing as $z$ is decreased from the
fully connected network, in both phantom and self-avoiding networks
[Fig.~\ref{fig:G}(a)]. Below the critical point $z_{\mathrm{c}}$ we
find that the shear modulus deviates from the zero-temperature
behavior, becoming non-zero for all finite temperatures. For
$z>z_{\mathrm{c}}$ the shear modulus is largely insensitive to
temperature, while for $z<z_{\mathrm{c}}$, $G$ depends strongly on
$T$. For high temperatures, the shear modulus becomes increasingly
insensitive to $z$ and deviates from the zero-temperature behavior at
increasingly high connectivities above $z_{\mathrm{c}}$, until
eventually, when $k_{B}T\sim k_{\mathrm{sp}}\ell_{0}^{2}$ (where
$k_{B}$ is the Boltzmann constant), the thermal energy of the system
is such that the network structure becomes unimportant.

The different regimes of the dependence of $G$ on $T$ can be seen in
Fig.~\ref{fig:G}(b). At high connectivities the shear modulus remains
almost constant as the temperature is increased, rising only as the
thermal energy $k_{B}T$ approaches the spring energy
$k_{\mathrm{sp}}\ell_{0}^{2}$. As we approach the critical point,
however, we find that the shear modulus, which will be $0$ at $T=0$,
shows an approximate $T^{1/2}$ dependence at low temperatures. This
anomalous temperature dependence is apparent over many orders of
magnitude, and in fact corresponds to the system becoming stiffer than
expected at low $T$ for ordinary entropic elasticity. As we increase
the temperature further, in the self-avoiding spring networks we see
this $T^{1/2}$ dependence give way to linear $T$ dependence, while in
the phantom spring networks we see a steeper $T$ dependence, although
it does not become linear. For $z<z_{\mathrm{c}}$ we find another
anomalous regime with $G\propto T^{\alpha}$, where $\alpha\simeq 0.8$,
at low temperature followed by linear $T$ dependence at high
temperatures in both phantom and self-avoiding networks. As we see
these anomalous regimes in both types of network, we conclude that
they are not driven by steric interactions, but instead by the random
network structure of these low $z$ value systems. Consistent with
this, if we remove bonds in such a way as to leave one-dimensional
chains of springs (i.e., chains with $z=2$) or honeycomb lattices
(with $z=3$) we find $G\propto T$ even at low temperatures, as one
would expect for ordinary entropic elasticity \cite{ref:SM}.

The observed shear moduli can be well described by a scaling form
analogous to that of the conductivity of a random resistor network
\cite{ref:Straley} that has also been successfully used to describe the
shear moduli of athermal spring and fiber networks
\cite{ref:Wyart,ref:Chase_bend}. For our system, this scaling Ansatz
is given by
\begin{equation}
\label{eq:G_scale}
G = k_{\mathrm{sp}}|\Delta z|^{a} \mathscr{F}\left(T^{*}|\Delta z|^{-b}\right),
\end{equation}
where $a$ and $b$ are constants, $\Delta z = z-z_{\mathrm{c}}$ and the
function $\mathscr{F}$ is dimensionless, as is its argument. We find
the best collapse of the data at low temperatures ($T^{*}=
k_{B}T/k_{\mathrm{sp}}\ell^{2}_{0}<10^{-5}$) for both the
self-avoiding and phantom networks using the exponents $a=1.4$ and
$b=2.8$, as shown in Fig.~\ref{fig:G}(c). This again demonstrates the
three low temperature regimes, with almost constant $G$ for
$z>z_{\mathrm{c}}$, $G$ scaling with $k_{\mathrm{sp}}T^{*0.8}$ $(\sim
k^{0.2}_{\mathrm{sp}}T^{0.8})$ for $z<z_{\mathrm{c}}$ and $G$ showing
$k_{\mathrm{sp}}T^{*1/2}$ $(\sim k^{1/2}_{\mathrm{sp}}T^{1/2})$
dependence as $\Delta z\rightarrow0$. We note that, similar to our
findings, a recent study of athermal fiber networks in two dimensions,
with both filament stretching described by $k_{\mathrm{sp}}$ and bond
bending described by stiffness $\kappa$, found that the shear modulus
scales with $k_{\mathrm{sp}}^{1/2}\kappa^{1/2}$ at the critical
connectivity \cite{ref:Chase_bend}.

\begin{figure}[h]
  \begin{center}
    \includegraphics[height=1.0\columnwidth,angle=270]{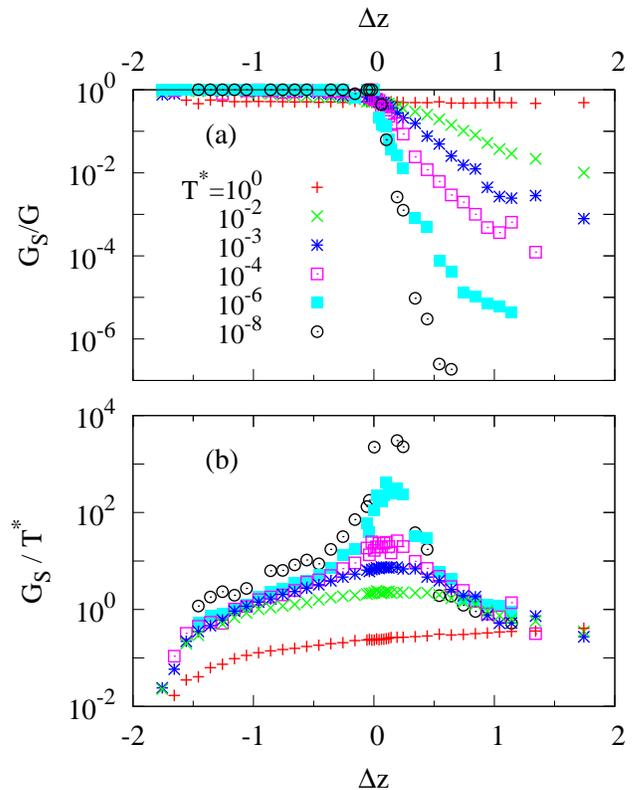}
    \caption{(a) The ratio $G_{S}/G$ as a function of $\Delta z$ for a
      phantom network at
      $T^{*}=k_{B}T/k_{\mathrm{sp}}\ell_{0}^{2}$. (b) $G_{S}/T^{*} =
      -k_{\mathrm{sp}}\ell^{2}_{0}(\partial^{2} {\cal S}/\partial \gamma^{2})/Ak_{B}$
      (units of $k_{\mathrm{sp}}$) as a function of $\Delta z$ for the
      same systems as above. Results are for $A=A_{0}$, $N=3600$ and
      $\ell_{0}=1$.}
    \label{fig:S}
  \end{center}
\end{figure}

The non-zero shear modulus we find below $z_{\mathrm{c}}$ can be shown
to be entropic in origin.  The shear modulus can be broken down into
its energetic and entropic parts as
\begin{eqnarray}
\label{eq:G_parts}
G &=& \frac{1}{A}\left(\frac{\partial^{2} {\cal U}}{\partial\gamma^{2}} - T\frac{\partial^{2} {\cal S}}{\partial\gamma^{2}}\right) \nonumber \\
&=& G_{E} + G_{S},
\end{eqnarray}
where ${\cal S}$ is the entropy, and both $G_{E}$ and $G_{S}$ can be
calculated during our simulation runs \cite{ref:SM}. We first show the
ratio $G_{S}/G$ versus $z$ for the phantom networks in
Fig.~\ref{fig:S}(a). At low temperature we see that $G_{S}/G$ rises
sharply as $z$ approaches $z_{\mathrm{c}}$ from above, before
saturating to $G_{S}/G\simeq1$ below $z_{\mathrm{c}}$, corresponding
to a dominant entropic contribution. For $z>z_{\mathrm{c}}$, the
energetic contribution $G_{E}$ dominates, although $G_{S}$ becomes
increasingly important at higher $T$.

\begin{figure}[h]
  \begin{center}
    \includegraphics[height=1.0\columnwidth,angle=270]{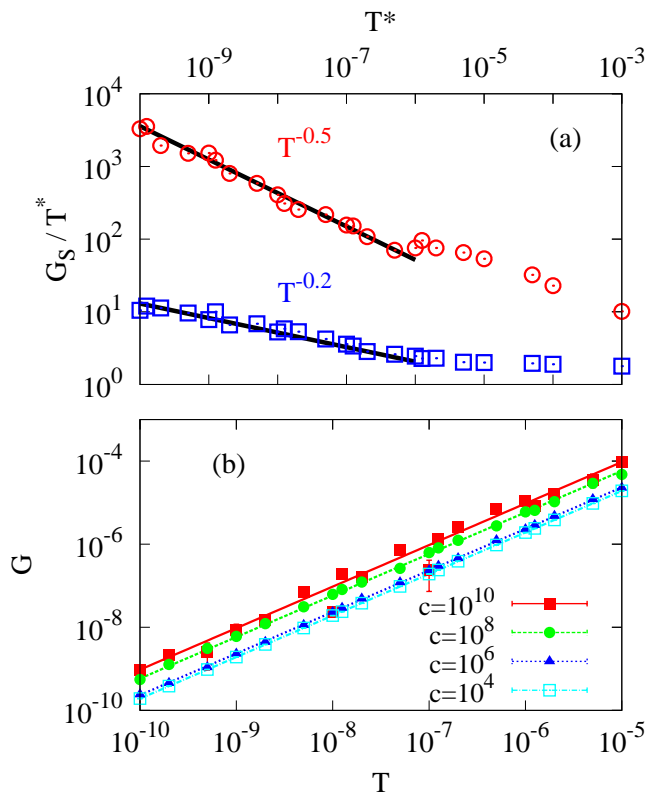}

    \caption{(a) $G_{S}/T^{*} = -k_{\mathrm{sp}}\ell^{2}_{0}(\partial^{2}
      {\cal S}/\partial \gamma^{2})/Ak_{B}$ against reduced
      temperature $T^{*}$ for phantom spring networks at area
      $A=A_{0}$ with $\ell_{0}=1$ and $z\simeq z_{\mathrm{c}} = 3.857$
      (red circles) and $z=3$ (blue squares). (b) Shear modulus $G$
      against temperature $T$ for phantom spring network with $z=3$
      and spring constant $k_{\mathrm{sp}} = cT$, where $c$ is a
      constant. Lines show linear fits}.
    \label{fig:S-2}
  \end{center}
\end{figure}

Figure~\ref{fig:S}(a) suggests that the behavior below the critical
point can be understood in terms of $G_{S}$ alone. Thus, when
considering the origins of the anomalous temperature dependence of the
shear modulus observed in Fig.~\ref{fig:G}, it is instructive to look
at the behavior of $\partial^{2} {\cal S}/\partial \gamma^{2}$ with
temperature and connectivity. From Eq.~(\ref{eq:G_parts}) it can be
seen that for pure entropic elasticity (where $G\propto T$) we should
see $\partial^{2} {\cal S}/\partial \gamma^{2} \propto T^{0}$. In
Fig.~\ref{fig:S}(b) we show $G_{S}/T^{*} =
-k_{\mathrm{sp}}\ell^{2}_{0}(\partial^{2} {\cal S}/\partial
\gamma^{2})/Ak_{B}$ against connectivity for a range of temperatures
in a system of phantom springs at constant area. As can be seen,
$G_{S}/T^{*}$ diverges at low temperatures as the critical point is
approached, both from above and below $z_{\mathrm{c}}$. In
Fig.~\ref{fig:S-2}(a) we show $G_{S}/T^{*}$ versus temperature. At the
critical point, we observe that $G_{S}/T^{*} \propto T^{-1/2}$ at low
temperatures. Similarly, for $z=3<z_{\mathrm{c}}$ we find that the low
temperature $G_{S}/T^{*} \propto T^{-0.2}$, before becoming
approximately constant at higher temperatures ($G_{S}/T^{*} \propto
T^{0}$). The high value of $\partial^{2} {\cal S}/\partial \gamma^{2}$
at low temperatures corresponds to the entropy decreasing more rapidly
as the system is sheared. As noted previously, for honeycomb-like
lattices and ideal chains we find ordinary entropic elasticity,
corresponding to $G_{S}/T^{*} \propto T^{0}$ throughout
\cite{ref:SM}. Hence, we conclude that the anomalous dependence of the
entropy on shear strain $\gamma$ at low temperatures arises from the
disordered nature of the network, leading to the anomalous temperature
dependence of the shear modulus. We note that we see qualitatively
similar behavior of $G_{S}/T^{*}$ with $T$ at low temperature for
self-avoiding networks, as one would expect from Fig.~\ref{fig:G}(b).

A possible origin of this anomalous temperature behavior in
sub-critical networks could be the internal stress $\sigma_{I}$ of the
network, which in the phantom networks arises from the resistance to
the tension the network is placed under in order to maintain its
area. This tension can be shown to be proportional to the temperature
\cite{ref:SM}. As such, at low temperatures the shear modulus can be
expected, on dimensional grounds, to scale as $G \propto
\sigma_{I}^{\alpha}k_{\mathrm{sp}}^{1-\alpha}$, which would appear as
$G \propto T^{\alpha}k_{\mathrm{sp}}^{1-\alpha}$ in our simulations. A
similar anomalous dependence on stress was found in athermal networks
with disordered molecular motors in
Ref.~\cite{ref:Misha_motor}. Interestingly, if one takes the spring
constant $k_{\mathrm{sp}}$ to be proportional to $T$, as would be
expected for freely-joined chains linking nodes, then pure entropic
elasticity would be recovered, with $G\propto T$ and $\partial^{2}
S/\partial \gamma^{2}\propto T^{0}$. However, if $k_{\mathrm{sp}}=
cT$, where $c$ is a constant, it follows from Fig.~\ref{fig:S-2}(a)
that the gradient of $G$ with $T$ would depend on the value of $c$. In
Fig.~\ref{fig:S-2}(b) we show the shear modulus against temperature
for networks with $z=3$ and $k_{\mathrm{sp}}= cT$, using a range of
$c$ values. Though all the systems show linear $T$ dependence, we do
see that as $c$ decreases, the shear modulus becomes smaller, until
$c\lesssim10^{5}$, where the results converge.

Our results demonstrate that there are two distinct regimes with
anomalous temperature dependence of the shear modulus, as illustrated
in Fig.~\ref{fig:PD}. In both cases, the dependence on $T$ is
\emph{sublinear}. Thus, at low temperatures, this corresponds to an
anomalously \emph{large} effect of thermal fluctuations. The natural
energy scale in our model is the spring energy
$k_{\mathrm{sp}}\ell_{0}^{2}$, which can easily be much larger than
the thermal energy, even at room temperature. For protein biopolymers,
for instance, it is expected that $k_{\mathrm{sp}}\simeq
Ed^{2}/\ell_0$, where the diameter $d$ is of the order of nanometers
and the Young's modulus $E$ can be as large as $1$ GPa
\cite{ref:Howard,ref:Pablo}, and hence the spring energy for a segment
of length $\ell_0\simeq100$nm can be more than $10^{6}$ times larger
than $k_{B}T$ at room temperature \cite{ref:note1}. Hence, for such
systems, reduced temperatures $T^{*}$ in the range $\lesssim10^{-6}$
can be relevant and network-level thermal fluctuations can be much
larger than expected based on naive entropic estimates. Importantly,
such network-level fluctuations are almost always ignored in prior
fiber network models and simulations, where either purely mechanical
models \cite{ref:Chase_bend,ref:Wyart,ref:Head,ref:Wilhelm,ref:Onck},
or hybrid mechanical models that include only single-filament
fluctuations \cite{ref:HuismanPRE,ref:HuismanPRL} have been
used. Finally, it is interesting to note that our phase diagram in
Fig.~\ref{fig:PD} is reminiscent of other systems with
zero-temperature critical behavior, such as quantum-critical points
\cite{ref:Sachdev,ref:Coleman}. As in such systems, in which the
critical point is also governed by fluctuations other than thermal, we
find a broad critical regime that fans out and extends for
temperatures potentially far above $T=0$.

\begin{acknowledgments}
This work was supported in part by a research programme of the 
Foundation for Fundamental Research on Matter (FOM, part of NWO).
\end{acknowledgments}


\begin{thebibliography}{10}
\bibitem{ref:note1}%
  \BibitemOpen
  \bibinfo {note} {In the case of semiflexible polymers, thermal fluctuations
  can lead to a reduced longitudinal spring constant, as well as a different
  $T$ dependence \cite{ref:Storm,ref:Odijk,ref:Mac_PRL}. This mechanical
  estimate is valid, however, for biopolymers of stiffness comparable to or
  greater than that of F-actin. Taking the spring constant to be thermal rather
  than mechanical, using $k_{\mathrm{sp}}=90 k_{B}T
  \ell^{2}_{\mathrm{p}}/\ell^{4}_{\mathrm{c}}$, a value of $T^{*}<10^{-6}$ is again
  found at room temperature (using values for the persistence length of
  $\ell_{\mathrm{p}}\sim10\mathrm{\mu m}$ and the contour length of
  $\ell_{\mathrm{c}}\sim100\mathrm{n m}$) \cite{ref:Head_PRE}.}%
  \bibAnnoteFile{Stop}{ref:note1}%
\end{thebibliography}
\end{document}


\widetext
\title{Supplementary material for: Fluctuation-stabilized marginal networks and anomalous entropic
  elasticity}

\author{M. Dennison$^{1}$} 
\author{M. Sheinman$^{1}$} 
\author{C. Storm$^{2}$} 
\author{F.C. MacKintosh$^{1}$} 

\affiliation{$^{1}$Department of Physics and Astronomy,
  Vrije University, De Boelelaan 1081, 1081 HV Amsterdam, The
  Netherlands}
\affiliation{$^{2}$Department of Applied Physics and Institute for Complex Molecular Systems, Eindhoven University of Technology,
  P.O. Box 513, NL-5600 MB Eindhoven, The Netherlands}

\begin{abstract}

In this supplementary material, we present the method used to
calculate the shear modulus, its energetic and entropic contributions
and the pressure of model spring networks. We show that the pressure
of a sub-critical network ($z<z_{c}$) held at constant area is
proportional to the temperature. Finally, we show that the linear
shear modulus of both a honeycomb lattice and a freely-jointed chain
are linear in temperature.
\end{abstract}

\maketitle

\section{Method}
\label{sec:Method}
We perform Monte Carlo (MC) simulations on systems of two-dimensional
spring networks. The networks consist of $N=n^{2}$ nodes arranged on a
triangular lattice and connected by $N_{\mathrm{sp}}=zN/2$ springs,
where $z$ is the connectivity ($z=6$ for the fully connected
network). We consider two cases: in one case we keep the system area
$A$ fixed (using $NVT$ simulations, where $T$ is the temperature and
we note that in our case the volume $V$ corresponds to the area $A$)
and treat the springs as `phantom' (i.e. we ignore steric
interactions, and hence the springs are allowed to overlap). In the
other we fix the system pressure $P$ (using $NPT$ simulations, and
hence the system area is allowed to change) and prevent the springs
from overlapping (i.e the springs self-avoiding). In both cases the
spring energy is given by

\begin{eqnarray}
\label{eq:E_sp}
E_{\mathrm{sp}}(\ell_{ij}) &=&
\frac{k_{\mathrm{sp}}}{2}\left(\ell_{ij}-\ell_{0}\right)^{2}\nonumber\\ &=&\frac{k_{\mathrm{sp}}}{2}\left(\sqrt{x_{ij}^{2}+
  y_{ij}^{2}}-\ell_{0}\right)^{2},
\end{eqnarray}
where $\ell_{ij}$ is the length of the spring that connects node $i$
to node $j$ (if $i$ and $j$ are connected by springs), $\ell_{0}$
represents the rest length and $k_{\mathrm{sp}}$ is the spring
constant. The total system energy ${\cal U}$ is then given as

\begin{equation}
\label{eq:U}
{\cal U} =\displaystyle\sum_{m=1}^{N_{\mathrm{sp}}}E_{sp,m} +
\displaystyle\sum_{m<m^{\prime}}^{N_{\mathrm{sp}}}V_{mm^{\prime}},
\end{equation}
where the first sum is over all springs $m$ and the second sum is the
interaction potential $V$ between all pairs of springs. For phantom
springs $V_{mm^{\prime}} = 0$, while for the self-avoiding springs
$V_{mm^{\prime}} = 0$ for non-overlapping pairs and $V_{mm^{\prime}} =
\infty$ for overlapping ones.

In order to lower the connectivity $z$ of a network, we set
$k_{\mathrm{sp}}=0$ for randomly chosen springs. For phantom networks
this is identical to removing springs, while for self-avoiding
networks setting $k_{\mathrm{sp}}=0$ has the advantage of
computational efficiency over removing the springs. For fully
connected triangular lattice networks the overlap test simply consists
of calculating the area of the triangles formed by the springs and
checking if it changes sign when a node is displaced. Springs with
$k_{\mathrm{sp}}=0$ will still contribute steric interactions which
allows for this overlap test to be used even at low connectivity, and
hence the nodes are essentially confined to a `cell' by the
surrounding springs.

The systems are allowed to relax at temperature $T$ in an MC
equilibration run of $\gtrsim 2\times10^{6}$ MC cycles. For the
self-avoiding networks we then perform a production run of $\gtrsim
4\times10^{6}$ MC cycles to calculate the average dimensions of the
simulation box. We finally fix the self-avoiding networks to their
average area and equillibrate them using $NVT$ simulations. 

In the phantom networks, where the area is fixed, we are effectively
applying a pressure on the network which will vary with the
temperature. We may calculate this pressure in a production run using
the expression for the virial pressure given as \cite{ref:Press}

\begin{equation}
\label{eq:pressure}
P = \frac{k_{B}T N}{A} + \frac{1}{2A}\displaystyle\sum_{m}^{N_{\mathrm{sp}}}
\left\langle f_{m}\ell_{m}\right\rangle ,
\end{equation}
where $k_{B}$ is the Boltzmann constant, $f_{m} =
k_{\mathrm{sp}}(\ell_{m} - \ell_{0})$ is the force of spring $m$ which
has length $\ell_{m}$ and $\langle \dots\rangle $ denotes the ensemble
average. Note that this expression is only valid for phantom spring
networks.

In order to shear the systems, we make use of Lees-Edwards boundary
conditions \cite{ref:Lees_Edwards} in our equilibrated systems. The
energy of springs that connect the nodes at the top of the box to
those at the bottom (via periodic boundary conditions) is modified to
become

\begin{equation}
\label{eq:E_sp_shear}
E_{\mathrm{sp}}(\ell_{ij}) =\frac{k_{\mathrm{sp}}}{2}\left(\sqrt{(x_{ij} +\gamma
  L_{y})^{2}+ y_{ij}^{2}}-\ell_{0}\right)^{2},
\end{equation}
where $L_{y}$ is the height of the simulation box, and $\gamma$ is the
shear strain. The shear modulus $G$ is given by

\begin{equation}
\label{eq:G_shear}
G =\frac{1}{A}\frac{\partial ^{2} {\cal F}}{\partial\gamma^{2}},
\end{equation}
where ${\cal F}$ is the free energy of the system. For $T=0$, ${\cal
  F}={\cal U}$ and as such $G$ is simply given by the second
derivative of the energy of the system divided by the area. As it is
not possible to directly calculate ${\cal F}$ from MC simulations we
instead calculate the shear stress $\sigma$ from which we can obtain
$G$. The method we use is given in Ref.~\cite{ref:Hoover}, and here we
outline its application to our system. We consider the difference in
free energy $\Delta {\cal F}$ between a sheared and unsheared system
($\gamma=0$) given as

\begin{eqnarray}
\label{eq:deltaF}
\Delta {\cal F} &=& -k_{B}T[\ln(Z_{\gamma}) - \ln(Z_{0})]\\ &=&
-k_{B}T\left[\ln\left(\sum_{k} e^{-\beta
    {\cal U}_{\gamma,k}}\right)-\ln\left(\sum_{k} e^{-\beta
    {\cal U}_{0,k}}\right)\right],\nonumber
\end{eqnarray}
where $Z$ is the partition function of the system, $\beta =
1/(k_{B}T)$, and the sum is over all states $k$. The energy of the
sheared system is ${\cal U}_{\gamma}$ and ${\cal U}_{0}$ is the energy
of the unsheared system. Taking the derivative of $\Delta {\cal F}$
with respect to $\gamma$ and dividing by the area $A$, we obtain the
shear stress as

\begin{eqnarray}
\label{eq:deltaF_2}
\sigma=\frac{1}{A}\frac{\partial {\cal F}}{\partial\gamma} &=& \frac{1}{A}\frac{\partial \Delta {\cal F}}{\partial\gamma}\nonumber\\ 
&=& \frac{1}{A}\frac{\displaystyle\sum_{k}\frac{\partial {\cal U}_{\gamma,k}}{\partial\gamma}e^{-\beta {\cal U}_{\gamma,k}}}{\displaystyle\sum_{k} e^{-\beta {\cal U}_{\gamma,k}}}\nonumber\\
&=&\frac{1}{A} \left\langle \frac{\partial {\cal U}_{\gamma}}{\partial\gamma}\right\rangle,
\end{eqnarray}
which is the ensemble average of the derivative of total system energy
with respect to $\gamma$. This is simply equal to the sum of the
derivatives of all spring and interaction energies

\begin{equation}
\label{eq:deltaF_3}
\sigma = \frac{1}{A}\displaystyle\sum^{N_{\mathrm{sp}}}_{m} \left\langle \frac{\partial E_{sp,m}}{\partial\gamma}\right\rangle + 
\frac{1}{A}\displaystyle\sum^{N_{\mathrm{sp}}}_{m<m^{\prime}} \left\langle \frac{\partial V_{mm^{\prime}}}{\partial\gamma}\right\rangle.
\end{equation}
We note that these derivatives are non-zero only for springs with the
energy given by Eq.~(\ref{eq:E_sp_shear}) (i.e. those crossing the
periodic boundary conditions). The derivative of
Eq.~(\ref{eq:E_sp_shear}) is given by

\begin{equation}
\label{eq:dE_sp_shear}
\frac{\partial E_{sp}}{\partial\gamma} =
\frac{k_{\mathrm{sp}}L_{y}\left(\sqrt{y_{ij}^{2} + (x_{ij}+\gamma
    L_{y})^{2}}-\ell_{0}\right)\left(x_{ij}+\gamma L_{y}\right)}{\sqrt{y_{ij}^{2} +
    (x_{ij}+\gamma L_{y})^{2}}}.
\end{equation}
For phantom spring networks the second term in Eq.~(\ref{eq:deltaF_3})
is zero. For self-avoiding spring networks, however, this term must be
calculated. We begin by noting that the x-coordinates of the `top' node
of springs crossing the boundary has been displaced to become

\begin{equation}
\label{eq:x_to_x}
x_{\gamma} = x + \gamma L_{y}.
\end{equation}
($x_{\gamma} = x$ for springs that do not cross the boundary, and
hence this coordinate has no $\gamma$ dependence). We also define a
value $f$, based on the overlap test we use, as

\begin{equation}
\label{eq:f_over}
f = x_{ij}y_{ik} - x_{ik}y_{ij},
\end{equation}
which is the cross product of vectors along the springs that connect
nodes $i$ and $j$ and nodes $i$ and $k$. As noted previously, our
overlap test is simply to test if the area of a triangle made up of
three springs connecting nodes $i$, $j$ and $k$ changes sign, which
will happen when $f$ changes sign. We define the initial spring
vectors such that all initial $f$ values are positive. The Boltzmann
factor of the hard-body interaction, $\exp{[-\beta V]}$, is therefore
a step function which takes the value $0$ for $f<0$ and $1$ for
$f>0$. We may write the derivative of the Boltzmann factor with
respect to $\gamma$ as

\begin{eqnarray}
\label{eq:d_Boltz1}
\frac{\partial \exp{[-\beta V]}}{\partial\gamma} &=& -\beta \frac{\partial V}{\partial\gamma}\exp{[-\beta V]}. \\
\label{eq:d_Boltz2}
\frac{\partial \exp{[-\beta V]}}{\partial\gamma} &=&  \frac{\partial \exp{[-\beta V]}}{\partial f}  \frac{\partial f}{\partial x_{\gamma}}  \frac{\partial x_{\gamma}}{\partial\gamma}.
\end{eqnarray}
Equating Eq.~(\ref{eq:d_Boltz1}) and Eq.~(\ref{eq:d_Boltz2}), and
noting that the derivative of a step function is a delta function, we
arrive at the following expression

\begin{equation}
\label{eq:d_V_d_gamma}
\frac{\partial V}{\partial\gamma} = -k_{B}T L_{y} \frac{\partial f}{\partial x_{\gamma}} \delta(f=0),
\end{equation}
which if inserted into Eq.~(\ref{eq:deltaF_3}) would involve
calculating the ensemble average of $\partial f/\partial x_{\gamma}$
for configurations of springs that are just in contact ($f=0$). During
a Monte Carlo simulation, sampling the contact value exactly would be
impossible. Instead, we follow the method used to approximate the
contact value of hard bodies (see e.g. Ref.~\cite{ref:P_W_spheroid}).

We calculate the average values of Eq.~(\ref{eq:dE_sp_shear}) and
Eq.~(\ref{eq:d_V_d_gamma}) during simulation runs for a range of
$\gamma$ values, from which we obtain $G$ simply as the derivative of
Eq.~(\ref{eq:deltaF_3})

\begin{equation}
\label{eq:G_shear_stress}
G =\frac{\partial\sigma}{\partial\gamma}.
\end{equation}

To calculate the entropic and energetic contributions to the shear
modulus, we first note that the free energy can be defined as

\begin{equation}
\label{eq:F_part}
{\cal F} = {\cal U} - T{\cal S},
\end{equation}
where ${\cal S}$ is the entropy. We can hence write the shear stress as

\begin{equation}
\label{eq:G_part}
\sigma =\frac{1}{A}\frac{\partial {\cal U}}{\partial\gamma} -
\frac{T}{A}\frac{\partial {\cal S}}{\partial\gamma},
\end{equation}
from which we define

\begin{eqnarray}
\label{eq:sigE_sigS}
\sigma_{E} &=& \frac{1}{A}\frac{\partial {\cal U}}{\partial\gamma }\nonumber\\
\sigma_{S} &=& - \frac{T}{A}\frac{\partial {\cal S}}{\partial\gamma },
\end{eqnarray}
and

\begin{eqnarray}
\label{eq:GE_GS}
G_{E} &=& \frac{\partial \sigma_{E}}{\partial\gamma }\nonumber\\
G_{S} &=& \frac{\partial \sigma_{S}}{\partial\gamma },
\end{eqnarray}
giving the shear modulus as $G=G_{E} + G_{S}$. The system energy
${\cal U}$ is calculated as the ensemble average energy during the
simulation run $\langle {\cal U}\rangle $, which is defined as

\begin{equation}
\label{eq:U_av}
\langle {\cal U}\rangle  =\frac{\displaystyle\sum_{k} {\cal U}_{\gamma,k} e^{-\beta {\cal U}_{\gamma,k}}}{\displaystyle\sum_{k}e^{-\beta {\cal U}_{\gamma,k}}}.
\end{equation}
By taking the derivative of this with respect to $\gamma$ we obtain

\begin{eqnarray}
\label{eq:dU_av}
\sigma_{E} &=& \frac{1}{A}\frac{\partial\langle {\cal U}\rangle }{\partial\gamma }\\
&=&\frac{1}{A}\left[\left\langle \frac{\partial {\cal U}_{\gamma}}{\partial\gamma}\right\rangle + 
\beta\left(\left\langle {\cal U}\right\rangle \left\langle\frac{\partial {\cal U}_{\gamma}}{\partial\gamma} \right\rangle   -
\left\langle {\cal U}\frac{\partial {\cal U}_{\gamma}}{\partial\gamma} \right\rangle \right)\right], \nonumber
\end{eqnarray}
and with Eq.~(\ref{eq:G_part}), Eq.~(\ref{eq:sigE_sigS}) and
Eq.~(\ref{eq:deltaF_2}) we can show

\begin{equation}
\label{eq:dS_av}
\sigma_{S} =  \frac{1}{A}\beta\left(
\left\langle {\cal U}\frac{\partial
  {\cal U}_{\gamma}}{\partial\gamma} \right\rangle -
\left\langle
{\cal U}\right\rangle \left\langle\frac{\partial {\cal U}_{\gamma}}{\partial\gamma}
\right\rangle \right).
\end{equation}
We use this together with Eq.~(\ref{eq:GE_GS}) to obtain the entropic
contribution to the shear modulus.

\section{Pressure of constant area system}
\label{sec:P}

\begin{figure}[!h]
  \begin{center}
      \includegraphics[height=0.7\columnwidth,angle=270]{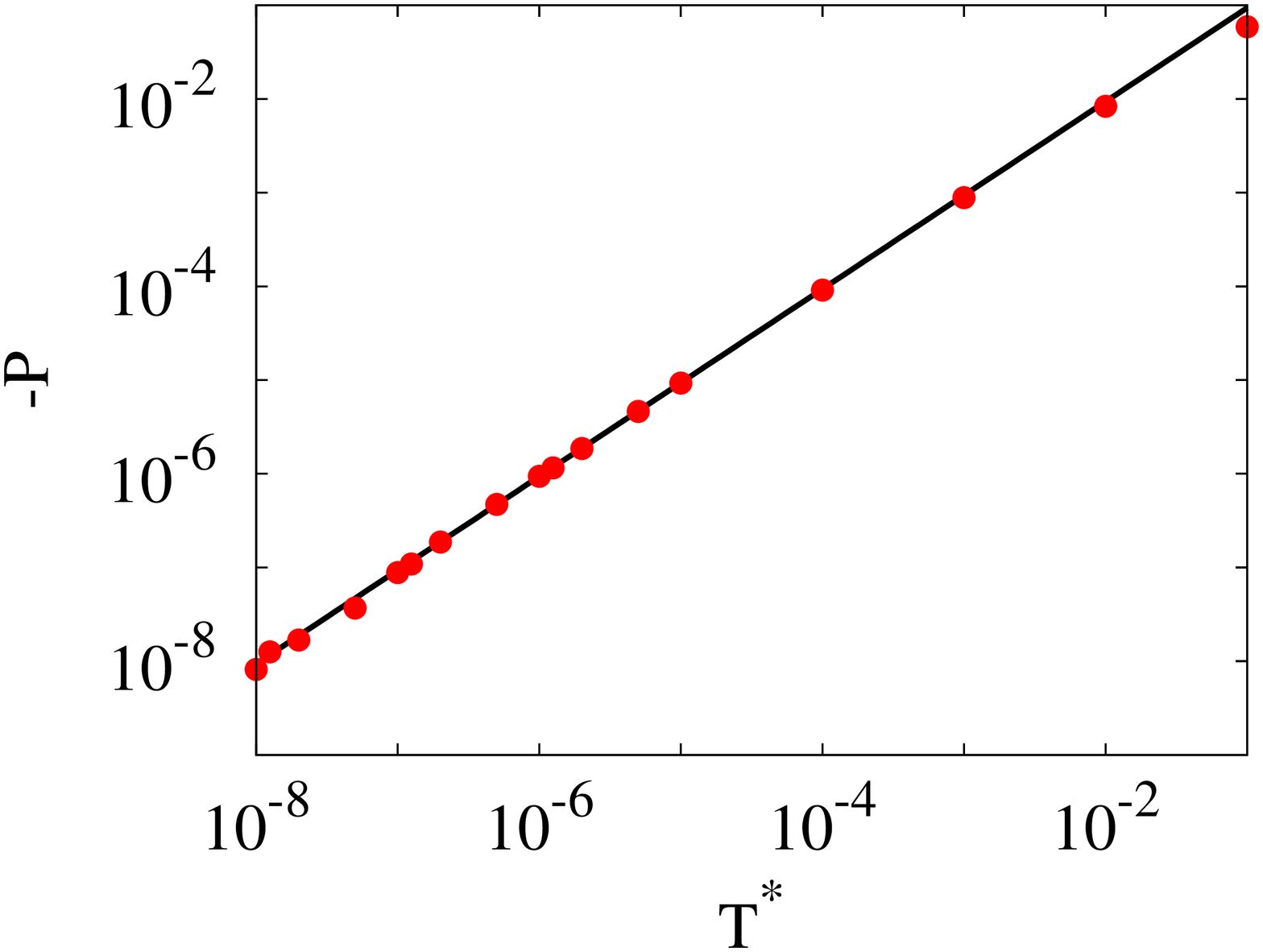}
    \caption{Pressure $P$ in units of $k_{\mathrm{sp}}$ of a phantom
      spring network with $z=3$ at constant area $A=A_{0}$ (where
      $A_{0}$ is the rest area of a fully connected network at $T=0$)
      as a function of reduced temperature $T^{*} = k_{B}T /
      k_{\mathrm{sp}}\ell_{0}^{2}$. Points are simulation data, line
      shows a linear fit.}
   \label{fig:press}
  \end{center}
\end{figure}

We calculate the pressure of a network using the virial pressure
equation given in Eq.~(\ref{eq:pressure}) for connectivity $z=3$ and
show the pressure $P$ as a function of temperature $T$ in
Fig.~\ref{fig:press}. As can be seen, we find that these networks are
under tension ($P<0$), with a magnitude that increases linearly with
the temperature.

\section{Shear modulus of ordered systems}
\label{sec:G}

\begin{figure}[!h]
  \begin{center}
    \includegraphics[width=0.6\columnwidth]{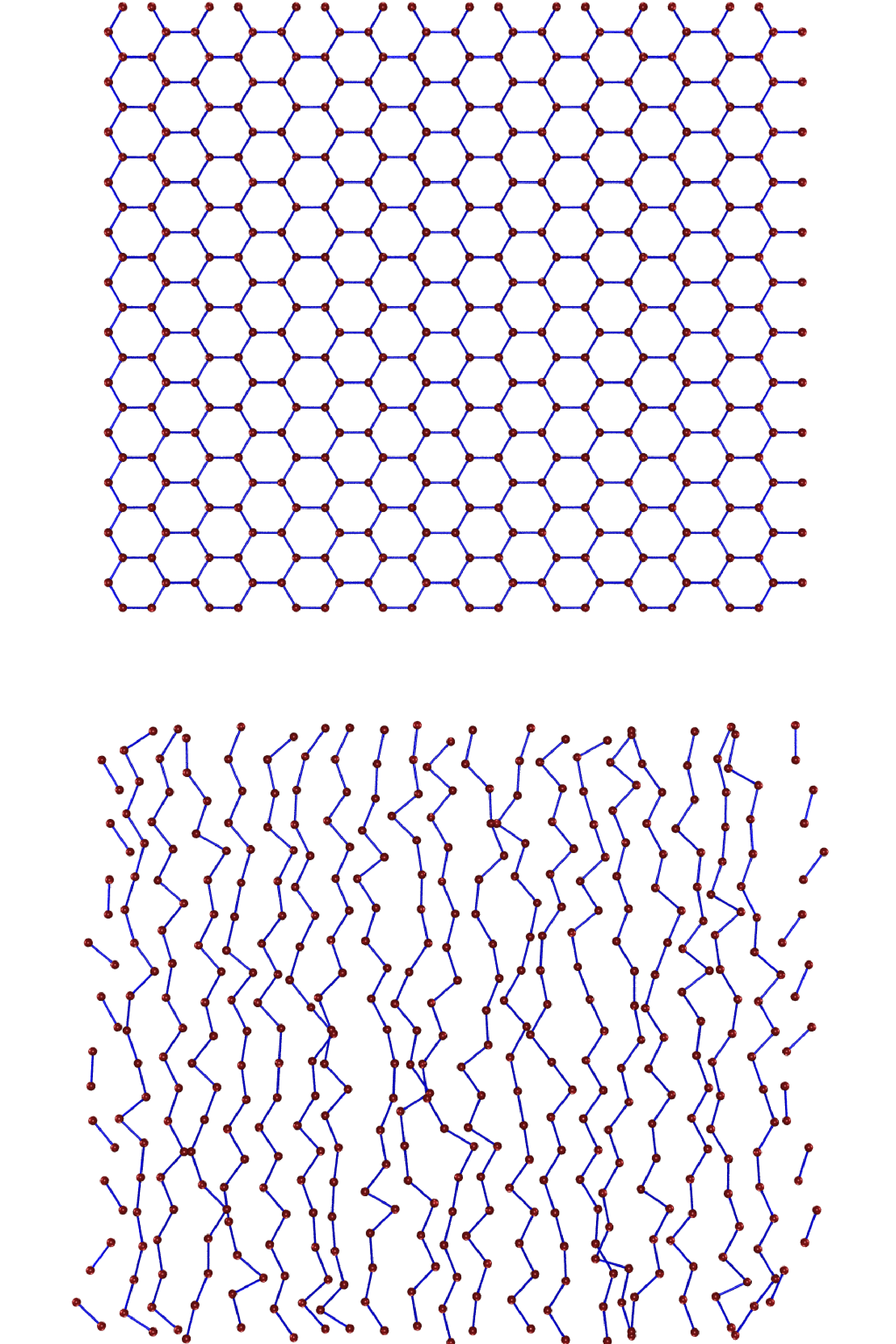}
    \caption{Example configurations of honeycomb lattice network (top)
      and freely-jointed chain (bottom). Blue lines are springs, red
      points are nodes.}
   \label{fig:FJC_HC}
  \end{center}
\end{figure}

\begin{figure}[!h]
  \begin{center}
      \includegraphics[height=0.7\columnwidth,angle=270]{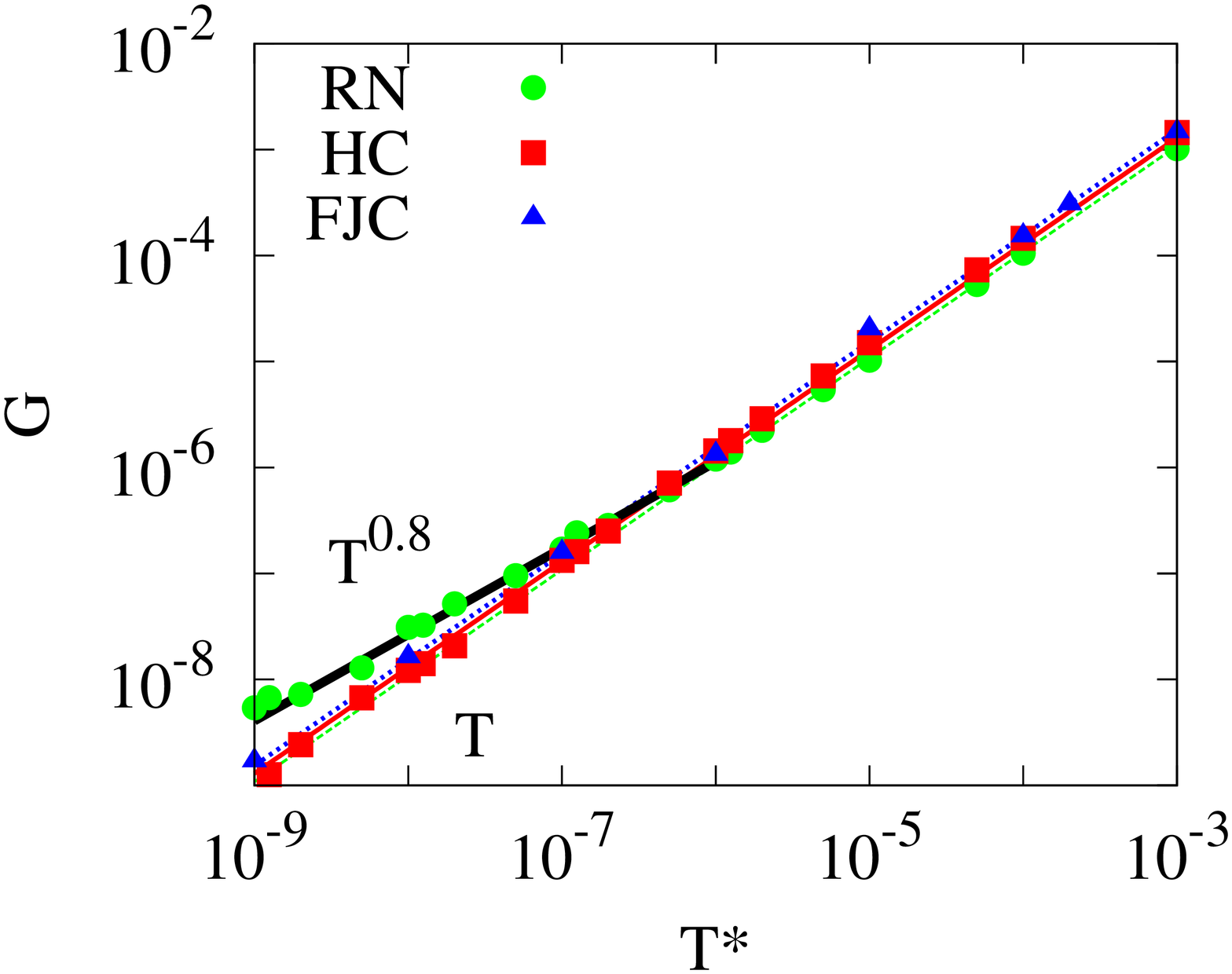}
    \caption{Shear modulus $G$ in units of $k_{\mathrm{sp}}$ of
      honeycomb lattice network (HC), freely-jointed chain (FJC) and
      randomly diluted triangular lattice network (RN) as a function
      of reduced temperature $T^{*} = k_{B}T /
      k_{\mathrm{sp}}\ell_{0}^{2}$ with area $A=0.9A_{0}$, where
      $A_{0}$ is the rest area of a fully connected triangular network
      at $T=0$. Points show simulation data, lines are to guide the
      eye with the labelled gradient.}
   \label{fig:G}
  \end{center}
\end{figure}

We calculate the linear shear modulus $G$ of both freely-jointed
chains and honeycomb lattice networks (example configurations of which
are shown in Fig.~\ref{fig:FJC_HC}) using the method given in
Section~\ref{sec:Method}. Both these networks are sub-critical
($z<z_{c}$) and each node in the networks has the same connectivity
($z=2$ for the freely-jointed chain, $z=3$ for the honeycomb lattice
networks). The behavior of the shear modulus $G$ of these systems with
temperature $T$, along with that of a randomly diluted triangular
lattice network, is shown in Fig.~\ref{fig:G}, where it can be seen
that $G$ shows linear $T$ dependence in both the honeycomb and freely
jointed chain systems, but not in the random network.

